\documentstyle[prl,aps,multicol,tighten,epsf]{revtex}

\begin{document}
\draft

\title{Streamer Propagation as a Pattern Formation Problem: 
Planar Fronts}
\author{Ute Ebert$^1$,  Wim van Saarloos$^1$ and Christiane Caroli$^2$}
\address{
$^1$Instituut--Lorentz, Universiteit Leiden, Postbus 9506, 2300 RA Leiden, 
the Netherlands,
       }
\address{
$^2$Universit\'e Paris VII, GPS Tour 23, 2 Place Jussieu, 75251 Paris Cedex 05,
France       }
\date{\today}
\maketitle

\begin{abstract} 
  Streamers often constitute the first stage of dielectric breakdown
  in strong electric fields: a nonlinear ionization wave transforms a
  non-ionized medium into a weakly ionized nonequilibrium plasma. New
  understanding of this old phenomenon can be gained through modern
  concepts of (interfacial) pattern formation. As a first step towards
  an effective interface description, we determine the front width,
  solve the selection problem for planar fronts and calculate their
  properties.  Our results are in good agreement with many features of
  recent three-dimensional numerical simulations.
\end{abstract}

\begin{multicols}{2}

  Transient discharges occur in various forms \cite{Gal}, e.g., as
  leaders in spark formation or as streamers in $ac$ silent discharges
  \cite{silent}. A common feature is the creation of a nonequilibrium
  plasma through the propagation of a nonlinear ionization wave into a
  previously non-ionized region. Although it is well known that,
  depending on the polarity of the field, discharge patterns on a
  larger scale may either be fractal \cite{pietronero}, or form more
  regular non-fractal patterns \cite{maan}, ionization fronts do not
  seem to have been analyzed before as a pattern forming system on
  scales resolving their internal structure. While the idea of a shock
  front or a thin ionization sheet has been formulated in the
  literature on streamers in the seventies \cite{fowler}, the
  analytical treatment then frequently was based on ad-hoc assumptions
  and on equilibrium concepts, e.g., on the assumption that the high 
  electric field would rise the electron temperature and that subsequent
  ionization would be thermal. In the last ten years, models
  incorporating nonequilibrium impact ionization of neutral molecules
  by free electrons have been investigated both numerically
  \cite{DW,Vit} and analytically \cite{Dya}. Fig.\ 1$(a)$ shows a snapshot
  from a numerical study by Vitello {\em et al.} \cite{Vit} of the
  streamer equations, Eqs.\ (\ref{1})-(\ref{4}) below. Here,
  the evolution of the electron and ion densities between two planar
  electrodes with distance 0.5 cm and voltage difference 25 kV is
  integrated forward in time for parameter values describing N$_2$
  under normal conditions. At time $t=0$, the electron density was
  taken nonzero only in a small localized region near the upper
  negative electrode.  The figure shows the electron density 5.5 ns
  later. Each contour line indicates the increase of the electron
  density by a decade. The lines enclose a finger-like region (the
  body of the streamer), consisting of a non-equilibrium plasma; this
  region rapidly expands downwards towards the anode. In the region
  outside, the gas is essentially non-ionized.  The fact that the
  contour lines in the figure are very closely and about equidistantly
  spaced, illustrates that the electron density within a zone of the
  order of a few $\mu$m grows about exponentially by a factor of about
  $10^{10}$. Since the total charge density is negligible before as
  well as behind the front, streamer dynamics can be viewed as the
  propagation of a thin charged ionization sheet separating a
  non-ionized high field region from an ionized electrically screened
  region.

  Obvious and up to now unanswered questions are: What determines the 
  scale of the pattern (e.g., the lateral width of the finger-like 
  region), its velocity, the field-enhancement near the tip, and what 
  effect do the boundary and initial conditions have?
  Triggered by the observation of interface-like profiles in the 
  simulations \cite{DW,Vit} and by the fact that these are precisely 
  the questions that are studied in the field of interfacial pattern
  formation for, e.g., dendrites and viscous fingers \cite{reviews}, 
  we show here that pattern formation concepts provide a systematic 
  route to unravel precisely these aspects:

  {\em (i)} For planar fronts, we trace the ``great defect'',
  ``the inability of the theory to determine a value for the wave speed''
  [5(c)], to the fact that streamers are an example of 
  {\em front propagation into unstable states} in virtually all models 
  analyzed \cite{fowler,Dya}. For such problems, it is
  well-known that the velocity cannot be obtained just by analyzing
  uniformly translating fronts using standard methods. In the field of
  pattern formation, the mechanism of {\em dynamical front selection}
  has been understood in the last decade \cite{benjacob,vs}. In this
  paper, we show that this allows us to derive all essential
  properties of planar fronts for the model of the recent simulations
  \cite{DW,Vit}.

  {\em (ii)} Clearly, an analysis of planar streamer fronts does not
  suffice to explain the {\em global} questions of pattern formation
  posed above, such as the field enhancement in front of the streamer 
  head or the radius of curvature of the tip. However, both the 
  simulations \cite{DW,Vit} and our analysis show that the
  propagating charge sheet is only a few $\mu$m thick, while the tip 
  radius and the electrode spacing are of order mm or more.
  This separation of scales, that makes simulations so demanding, 
  can be made into an analytical tool. Much of our present knowledge 
  about similar problems like combustion fronts \cite{comb}, 
  thermal plumes \cite{plumes} and chemical waves \cite{chemwaves}, 
  etc., is based on an {\em effective interface description}. Such a 
  physically appealing formulation can be systematically derived in
  a matched asymptotic expansion to lowest order in the ratio 
  $\ell_{in}/ \ell_{out}$, where the inner length scale $\ell_{in}$ is 
  the thickness of the front (here the thickness of the charge sheet), 
  and $\ell_{out}$ the scale of the pattern, e.g., the tip radius.
  In the effective interface approach that we propose for streamers, 
  the charge sheet can be viewed as a {\em weakly curved} locally 
  almost planar front, since the thickness of the charge sheet is much 
  smaller than its radius of curvature. Like in the
  other problems, the importance of our planar front analysis
  therefore lies in the fact that, apart from curvature corrections,
  it provides a complete solution of the so-called inner problem. 

  {\em (iii)} In the non-ionized region outside the streamer, the
  electrical potential $\Phi$ obeys the Laplace equation, $\nabla^2
  \Phi = 0$. Moreover, our analysis shows that the normal velocity of
  a negatively charged planar streamer front ($v^*$ below) is a
  weakly nonlinear function of the field $E^+= - \nabla \Phi$ just
  ahead of it. Both features are reminiscent of the equations for
  other interfacial pattern forming problems like dendrites --- e.g.,
  the enhanced diffusion in front of a dendrite tip is analogous to
  the field enhancement in front of a streamer.
  Streamers will therefore be amenable to the
  same type of analysis \cite{reviews,comb,plumes}.
Physically, we expect that the interface equations will take
the form of a conservation equation for a charge sheet
(involving transport terms along the sheet, a stretch term
due to interface curvature and a term associated with charge 
transport from the plasma behind), supplemented with an
equation for the front speed that includes curvature corrections,
and an equation for the degree of ionization created by the front
which is not determined by any conservation law. 
The derivation of the appropriate equations is left to the future,
as the analysis is far from trivial due to the coupling to the
dynamics of the plasma, the fact that the electric field is
typically not normal to the front, and the fact, that in this
fully nonequilibrium situation, the curvature corrections do not 
follow from simple thermodynamic considerations.

  We now sketch our analysis \cite{ebert} of planar fronts in the
  streamer model equations \cite{DW,Vit,Dya} that also underly
  Fig.~1$(a)$. The electron and ion densities $n_e$, $n_+$, 
  and the electric field ${\cal E}$ obey the balance equations
\begin{eqnarray}
\label{1}
\partial_t\;n_e \;+\; \nabla_{\bf R}\cdot{\bf j}_e
&=& |n_e \mu_e {\cal E}| \;\alpha_0
\;\mbox{e}^{-E_0/|{\cal E}|}~,
\\
\label{2}
\partial_t\;n_+ \;+\; \nabla_{\bf R}\cdot{\bf j}_+
&=& |n_e \mu_e {\cal E}| \;\alpha_0
\;\mbox{e}^{-E_0/|{\cal E}|}~,
\end{eqnarray}
and the Poisson equation
\begin{equation}
\label{3}
\nabla_{\bf R}\cdot {\cal E} = {{e}\over{\varepsilon_0}} \;(n_+ -n_e)
~.
\end{equation}
The electron and ion current densities ${\bf j}_e$ and ${\bf j}_+$ are
\begin{equation}
\label{4}
{\bf j}_e = - n_e \;\mu_e \;{\cal E} - D_e \;\nabla_{\bf R} \;n_e ~,
~~~
{\bf j}_+ = 0~,
\end{equation}
so that ${\bf j}_e$ is the sum of a drift and a diffusion term, while
the ion current ${\bf j}_+$ is neglected, since the ions are much less
mobile than the electrons.  The r.h.s.\ of Eqs.\ (1) and (2) is a
source term due to the ionization reaction: In high fields free
electrons can generate free electrons and ions by impact on neutral
molecules. The source term is given by the magnitude of the electron
drift current times the target density times the effective ionization
cross section; the rate constant $\alpha_0$ has the dimension of an 
inverse length. The exponential function expresses, that only in 
high fields electrons have a nonnegligible probability to collect 
the ionization energy between collisions.

To identify the proper parameters for the behavior on the inner front scale,
we note that in the simulations the fields just ahead of the front are
of order of the threshold field $E_0=2\cdot 10^5$ V/cm in Eqs.\ 
(\ref{1})-(\ref{2}). The larger the rate parameter $\alpha_0$, the
more rapid the impact ionization will be, and the thinner the front
region. The natural length scale for the width of the front will indeed 
turn out to be $\alpha_0^{-1}$, which is about 2.3 $\mu$m in the
simulations \cite{DW,Vit}. As the drift velocity of electrons in a
field of order $E_0$ is $\mu_e E_0$, the natural timescale for the
motion of fronts is then $t_0=(\alpha_0 \mu_e E_0)^{-1}$ ($\approx
3\cdot 10^{-12}$ s in \cite{DW,Vit}) and the natural scale for
the charge density is $q_0=\varepsilon_0 \alpha_0 E_0$ ($\approx 4.7
\; 10^{14} \;e/ \mbox{cm}^3$ in \cite{DW,Vit}).

For analyzing planar fronts, we now introduce dimensionless variables, 
$x = X \alpha_0 $, $\tau=t/t_0$, $E={\cal E}/E_0$, the electron  
density $\sigma=en_e/ q_0$, and the total charge density 
$q=(n_+-n_e)e/q_0$. In these units, the {\em only} remaining
dimensionless parameter is the dimensionless diffusion coefficient 
$D= D_e \alpha_0 / \mu_e E_0$. In the simulations for N$_2$
\cite{DW,Vit}, this value is about 0.1; for typical gases, $D$ ranges
from 0.1 to 0.3 \cite{ebert}. In these variables, the charge
conservation equation becomes from (\ref{1}), (\ref{2}), (\ref{4}):
$\partial_\tau q+ \partial_x(\sigma E +D \partial_x \sigma)=0$. 
Upon combining this with the Poisson equation $\partial_x E=q$
and integrating, we obtain
\begin{equation}
\label{5}
\partial_\tau\;E
= \;-\;\sigma\;E\;-\;D\;\partial_x\; \sigma~.
\end{equation}
Here the integration constant is zero because on the inner time
and spatial scale the charge and electron densities vanish for
$x\to\infty$, while $E(x\to\infty) = E^+$ time independent.  
Eq.\ (\ref{5}) and the equation for the electron
density
\begin{equation}
\label{6}
\partial_\tau\;\sigma 
= \partial_x\;(\sigma \;E) \;+\; D \;\partial_x^2 \sigma \;+\;
\sigma \; |E|\; \mbox{e}^{-1/| E|} ~,
\end{equation}
together constitute the one-dimensional streamer equations. 
These equations have two important classes of steady state solutions: 
the ones with $\sigma=0$, $E^+$ arbitrary, correspond to the 
non-ionized state of the gas into which the front propagates. 
The ones with $\sigma$ constant (denoted $\sigma^-$) and $E=0$ 
correspond to the screened ionized state behind the front. 
It is straightforward to analyze the
linear stability of these states with Fourier modes of the form
e$^{\:\omega \tau +ikx}$. Physically, one expects the non-ionized
$(+)$ state to be {\em unstable}: Any small electron density
drifts in the field $E^+$ and gets amplified due to impact
ionization while the stabilization due to diffusion dominates
only at short wavelengths. The corresponding dispersion relation 
$\omega^+ = ikE^+ + |E^+|\;\mbox{e}^{-1/|E^+|} -Dk^2$ confirms the 
long wave length instability. It is easily 
checked that screening stabilizes the ionized $(-)$ states at all 
wavelengths, and that $\omega^- =-\;\sigma^--Dk^2$. 

Propagating streamer fronts are therefore an example of {\em front
propagation into an unstable state}. We thus follow the common path for
such problems \cite{benjacob,vs}: \\
{\em (a)} As usual, one can demonstrate the existence of a continuous 
family of uniformly translating front solutions of the form $\sigma(\xi)$ 
and $E(\xi)$ with $\xi=x-v\tau$, parametrized by the 
velocity $v$. This is done by formulating the equations for 
$\sigma(\xi)$ and $E(\xi)$ as a flow in
the phase space $(\sigma,E,\sigma')$ with $\xi$ playing the role of a
time-like variable. A front profile then corresponds to a trajectory
connecting one $(-)=(\sigma^-,0,0)$ fixed point with one
$(+)=(0,E^+,0)$ fixed point, and the existence and multiplicity of
these can be studied with counting arguments \cite{vs}.
The family of solutions can be obtained explicitly for $D=0$ 
by writing Eq.\ (\ref{5}) as $v\partial_\xi \ln |E|=\sigma$, by 
inserting this form into Eq.\ (\ref{6}) and integrating: we then get
$\sigma[E]={{v}/(v+E)}\int_{|E|}^{|E^+|} dx \; \exp(-1/x) $.
This determines the flow in phase space for $D=0$. \\
{\em (b)} Physically acceptable front solutions must satisfy 
the additional constraint that the number densities $n_e$ of 
electrons and $n_+$ of ions be positive, i.e., $\sigma(\xi) \ge0$ 
for all $\xi$. \\
{\em (c)} We can show that the
condition {\em (b)} entails a lower bound on the range of
velocities. More precisely, one can show \cite{ebert} that the
velocity of physically admissible front solutions obey
\begin{equation}
\label{7}
v \ge v_f = \mbox{max}~[v^*,v^\dagger ]  > 0,
\end{equation}
where $v^\dagger$ is the fastest {\em nonlinear front} \cite{vs} if it
exists. Nonlinear fronts correspond to  strongly heteroclinic orbits
in phase space: they reach the $(+)$ fixed point along the
eigendirection with the fastest contraction. 
The velocity
$v^* = -E^+ + 2\sqrt{D\;|E^+|\; \exp(-1/|E^+|)}$
is the value of the velocity $v$ below which the eigenvalues 
describing the flow close to the $(+)$ fixed point become complex, 
so that the $\sigma(\xi)$ profiles violate {\em (b)} as they
oscillate  around zero far ahead of the front. \\
{\em (d)} Existing knowledge of front propagation
\cite{benjacob,vs} leads us to conjecture the following
mechanism of front selection: {\em Fronts emerging from sufficiently
localized initial conditions} \cite{ft20} {\em converge asymptotically 
to the slowest physically acceptable front solution $v_f$ defined in}
(\ref{7}) \cite{footnote}. 

We have investigated the existence of nonlinear ($v^\dagger$) fronts
analytically and numerically and checked the above conjecture about
dynamical selection by direct numerical integration of Eqs.\ (\ref{5})
and (\ref{6}). Both qualitatively and quantitatively, our predictions
reflect the strong asymmetry between fronts moving parallel and
antiparallel to the field:

Fronts propagating parallel to
the electron drift, i.e., into a field $E^+<0$,  are {\em negatively
charged}. Numerically we find no $v^\dagger$ front solutions so that we
predict the selected front velocity to be always the value $v^*$ given
under {\em (c)}. Here diffusion and ionization help to raise the
front velocity to a value somewhat larger than the electron drift
velocity $-E^+$. The degree of ionization $\sigma^-$ behind the front
only weakly depends on $D$. The analytic result $\sigma^-=\sigma[E=0]$
for $D=0$ (see formula under {\em (a)}) \cite{Dya} is 
independent of $v$ and a good approximation for all physical values of 
$D$, as the lower panel of Fig.\ 1$(b)$ illustrates. Moreover, the values 
of $\sigma^-(E^+)$ extracted from the full $3D$ simulations of Vitello 
{\em et al.} \cite{Vit} (crosses) are close to the values we calculate 
for planar streamer fronts.

Fronts screening a field $E^+>0$ are {\em positively charged}.
They can propagate only, if diffusion overcomes the drift. 
As a result, for small $D$ propagating fronts are extremely 
steep and slow. The front velocity vanishes like $D$, while 
both, the spatial decay rate and the degree of ionization behind 
the front scale like $1/D$.  
In the limit $D\to0$ this singular behavior can be derived 
analytically \cite{ebert}. For general $D$, we have predicted
the front velocities $v_f(E^+,D)$ numerically.  
They are shown in Fig.\ 1$(b)$.

The numerical integration of the initial value problem fully supports
all our predictions on the asymptotically approached front for
sufficiently localized initial conditions. As an example, Fig.\ 2
shows the spatio-temporal development of electron density $(a)$
and field $(b)$ of an initial state with $E=-1$ and a small 
charge-neutral Gaussian ionization seed. The diffusion constant
is $D=0.1$, and the field far from the ionized region is held
constant. The ionized region initially grows exponentially and the
electrons drift with the field, till field screening in the middle
sets in.  Then a negative front emerges to the right and
asymptotically (after $\Delta t \approx 20$) approaches the $v^*$ 
(=1.38) front with $\sigma^- = 0.144$. The positive front on the left
initially recedes and then gets stuck by the combined action of drift
and screening.  This structure keeps slowly evolving in time, however,
till after a time of order $ 4000$, the predicted
positive front with $v^\dagger = 0.0146$ and $\sigma^- = 6.23$
emerges (not shown).

In summary, we have solved the planar streamer front problem. Based on
these results, we advocate that one should understand streamer
dynamics as a two-scale problem: on the inner scale, we have a moving
ionization sheet, whose thickness $\approx 10 \mu$m is set by the
ionization length $1/\alpha_0$. This interface plays the role of a
free boundary for the outer dynamics, whose scale is set by the global
geometry. {\em It is on this scale, that the pattern formation problem
  should be studied}. The similarity with other well-known interfacial
pattern forming problems (a Laplace equation for the potential $\Phi$
in the non-ionized region with, apart from curvature corrections, a
normal front velocity a function of $\nabla \Phi$), gives us
confidence that properties of streamer patterns like field enhancement
at the tip, velocity and tip radius, can be obtained in an analogous
way \cite{reviews} by including the curvature corrections in the 
resulting effective interface equations.

We thank M.~van Hecke and F.L.J.~Vos for help with the
graphics and P.A.~Vitello for providing Fig.\ 1$(a)$. 
UE is supported by the Dutch Science Foundation NWO.

\begin{figure}[h]
\setlength{\unitlength}{1cm}
\begin{picture}(8,6.5)
\epsfxsize=6.5cm
\end{picture}
\begin{center}
\begin{minipage}{8cm}
\small FIG.\ 1.
  {\em (a)} Electron density profile in a negative streamer from
  the 3D cylindersymmetric numerical simulations \cite{Vit} of Eqs.\ 
  (1)-(4). Courtesy of P.A.~Vitello. 
  {\em (b)} Our predictions for planar fronts. 
  Upper panel: $v^\dagger/D$ (solid) and $v^*/D$ (dashed lines) as a 
  function of $D$ for positive fronts and for $E^+ =$ 0.4, 0.6, 0.8, 
  1.0, 1.2, and 1.4, from bottom to top.
  Lower panel: Electron density $\sigma^- =n_e e/q_0$
  behind a negative front as a function of the field $E^+$ before the
  front for $D=$ 0 (solid line), 1 (dashes), 3 (dots).  Crosses:
  values of $\sigma^-(E^+)$ on the symmetry axis in the 3D simulations
  \cite{Vit} at times 4.75 ns and 5.5 ns, with $E^+$ the value of the
  outer field extrapolated towards the tip, in accord with the
  asymptotic matching prescription. 
\end{minipage}
\end{center}
\end{figure}

\begin{figure}[h]
\setlength{\unitlength}{1cm}
\begin{picture}(8,6.5)
\epsfxsize=6.5cm
\end{picture}
\begin{center}
\begin{minipage}{8cm}
\small FIG.\ 2.
  Numerical integration of Eqs.\ (5) - (6) for $D=0.1$ in a 
  constant background field $E=-1$. Initial state at $t=0$: lowest
  line. Each new line corresponds to a time step $\Delta t =5$ and the
  upper line to $t=100$. $(a)$ electron density, $(b)$ electric field.
\end{minipage}
\end{center}
\end{figure}

\end{multicols}

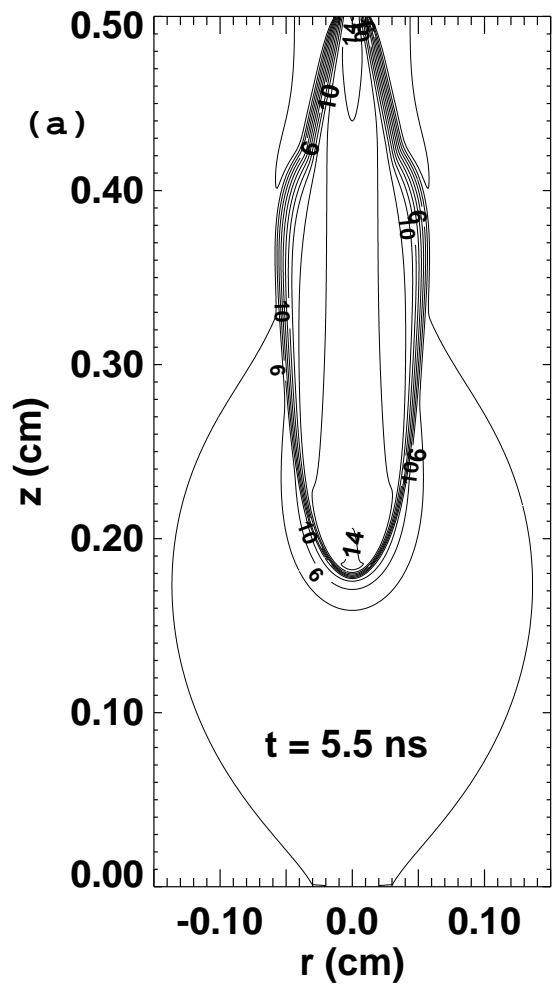
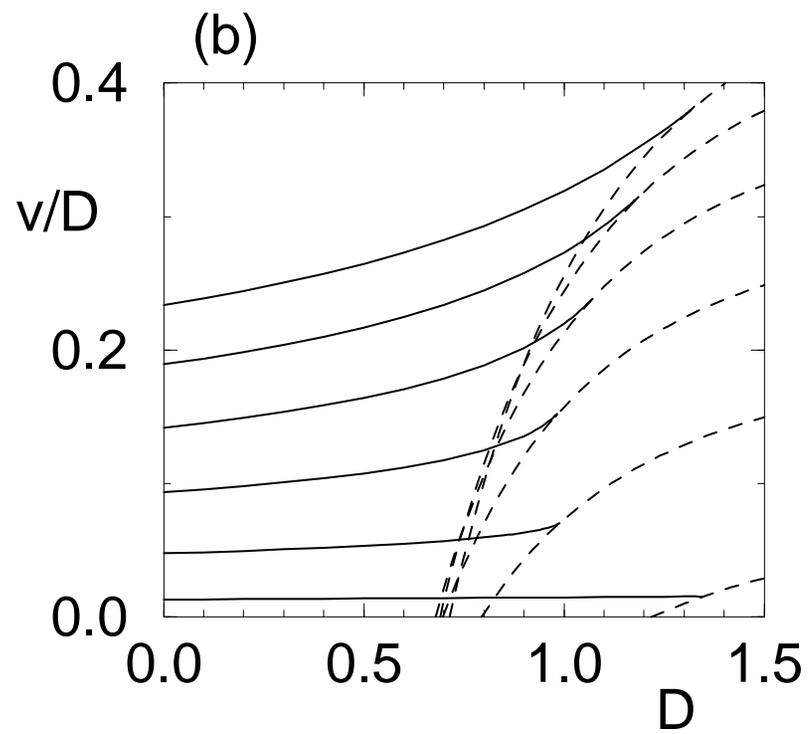
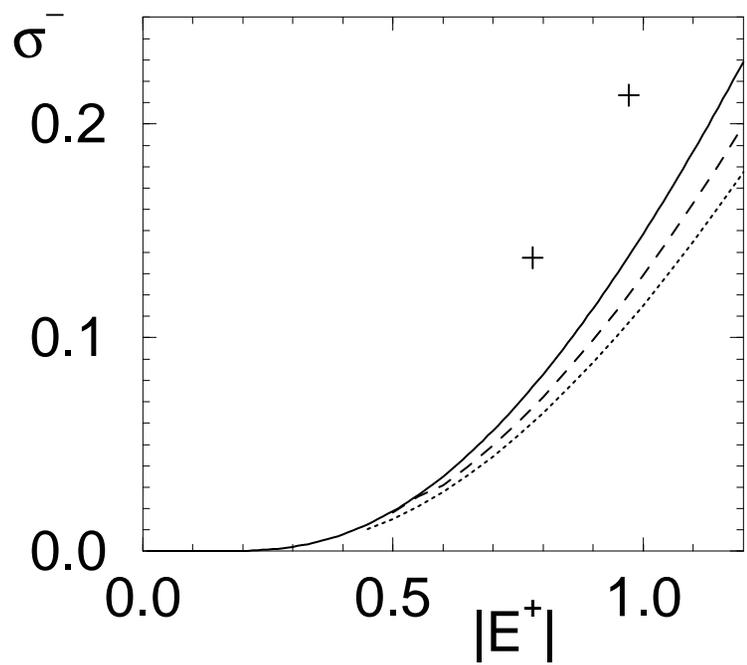
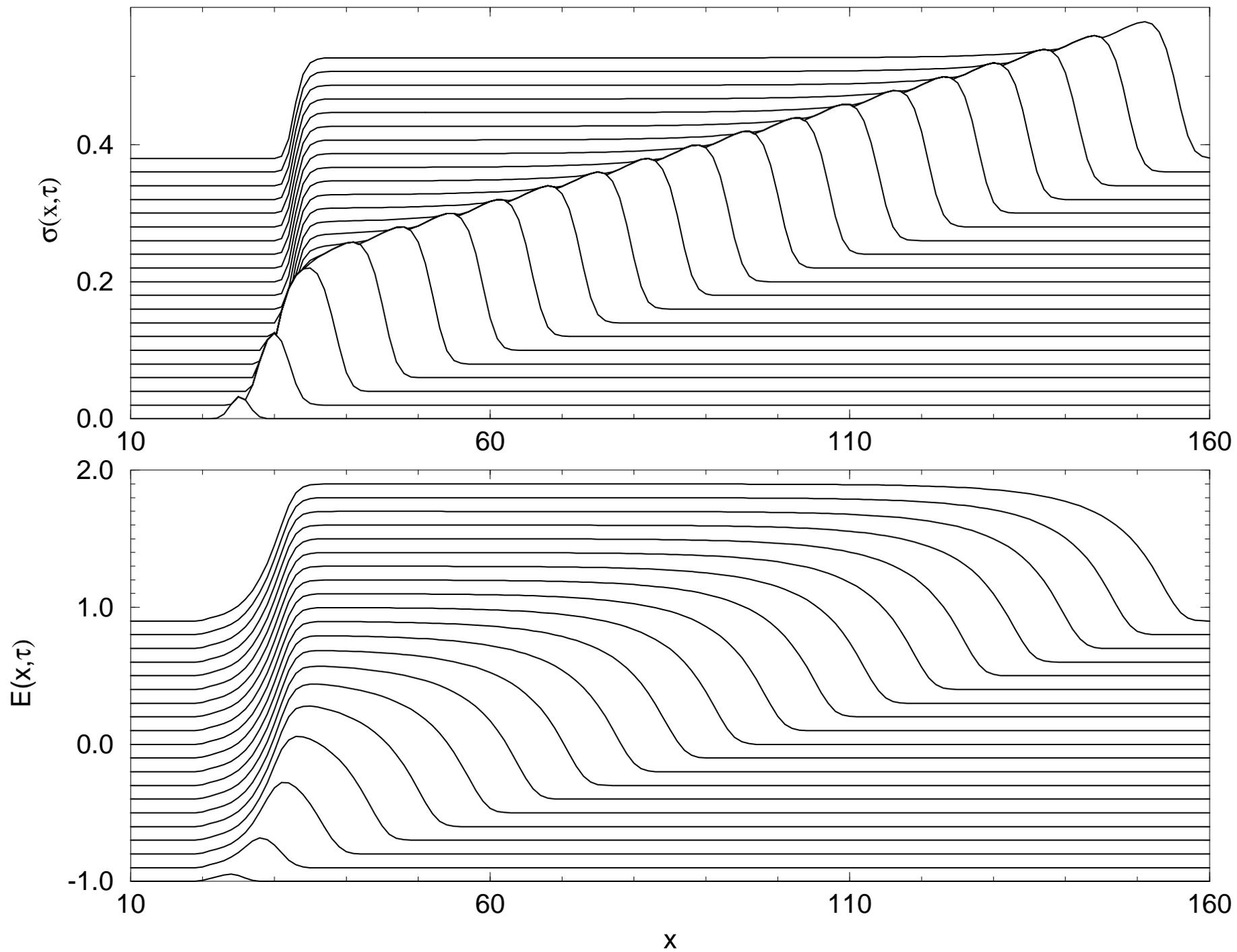
\end{document}